\documentstyle[epsf,12pt]{article}
\newcommand{\newsection}{    
\setcounter{equation}{0}
\section}
\def\appendix#1{
\addtocounter{section}{1}
\setcounter{equation}{0}
\renewcommand{\thesection}{\Alph{section}}
\sectition*{Appendix \thesection\protect\indent #1}
\addcontentsline{toc}{section}{Appendix \thesection\ \ \ #1}
}

\newcommand{\eq}[1]{Eq.~(\ref{#1})}

\newcommand{\beq}{\begin{equation}}
\newcommand{\eeq}{\end{equation}}
\newcommand{\bea}{\begin{eqnarray}}
\newcommand{\eea}{\end{eqnarray}}

\newcommand{\om}{\omega}

\newcommand{\tr}{{\,\rm tr}\:}

\newcommand{\ee}[1]{{\rm e}^{#1}\,}

\newcommand{\moy}[1]{{\left<{#1}\right>}}
\newcommand{\td}[1]{{\tilde{#1}}}

\newcommand{\ca}[1]{{#1}}
\newcommand{\cs}[1]{}

\title{The Potts-$q$ random matrix model : loop equations,
critical exponents, and rational case }
\begin{document}
\begin{titlepage}

\hfill
\begin{minipage}[t]{3.cm}
{\mbox{}\\ Saclay T99/060\\}
\end{minipage}
\\[1.cm]

\begin{center}
{\bf{ \LARGE The Potts-$q$ random matrix model : loop equations,
critical exponents, and rational case}}

\bigskip

{\large B. Eynard \footnote{eynard@theory.physics.ubc.ca}\, \dag \, \ddag \
 and G. Bonnet
\footnote{gabonnet@spht.saclay.cea.fr} \, \ddag}
\\
\dag \ Department of Physics and Astronomy, University of British Columbia
6224 Agricultural Road, Vancouver, British Columbia, V6T 1Z1\\
\ddag \ CEA/Saclay, Service de Physique Th\'eorique,  F-91191, Gif-sur-Yvette Cedex, 
France\\
\end{center}
\bigskip

\begin{abstract}

In this article, we study the $q$-state Potts random matrix
 models \cs{(in short, Potts-$q$)} extended to branched polymers, 
by the equations of motion method. We obtain \cs{non-trivial} \ca{a set of} loop equations
 valid for any arbitrary value of $q$.
We show that, for $q=2 -2 \cos {l \over r} \pi$ ($l$, $r$ mutually prime
integers with $l < r$ ), the resolvent satisfies an algebraic
equation of degree $2 r -1$ if $l+r$ \ca{is} odd and $r-1$ if $l+r$ \ca{is} even.
This generalizes the \cs{so far only known} \ca{presently-known} cases of $q=1,2,3$.
We then derive for any $0 \leq q \leq 4$
the Potts-$q$ critical exponents and string susceptibility.

\end{abstract}

\end{titlepage}%

\topmargin 0pt
\oddsidemargin 5mm
\headheight 0pt
\headsep 0pt
\topskip 9mm

\newsection{Introduction}

Random matrix models \cite{houches94} have proven to be a powerful mathematical
tool for the study of statistical physics systems on a fluctuating 
two dimensional lattice \cite{BIPZ}. In particular, the Potts model 
\cite{pottsplat} on a random lattice \cite{pottsrand}, which was first
partially solved by Daul \cite{daul}, has \cs{known}\ca{received} a recent renewed interest 
\cite{paul,GB} due to new approaches \cs{of}\ca{to} the problem.
It is a $q$-matrices model where all the
matrices are coupled to each other, which prevents \ca{one} from
using the formula \cite{IZ,HCh} to integrate
out the relative angles between the matrices (they no longer are
independent variables) and deal with the eigenvalues only.
In this paper we use the equations of motion method which does not 
involve integration over angular variables.
We obtain non-trivial loop equations relating even moments to odd moments
of a single matrix $M_i$, and we show how to extend these relations to the 
case of Potts-$q$ plus branched polymers (gluing of surfaces).
Such relations (which could not be obtained by previous methods 
\cite{daul,paul}) are needed to apply the renormalization group 
method \cite{RG} to Potts-$q$ models with added branching interactions, 
which is hoped to provide an understanding of the $c=1$ ($q=4$) barrier.

We  obtain an $O(n)$-like equation,
the solution of which is known \cite{Oden} and involves elliptical
functions \cite{elli}.
What was the resolvent in the $O(n)$ model, however, is now (up to some
transformations) the functional inverse of the Potts-$q$ resolvent,
 \cs{and}\ca{with} $n$  replaced by $2-q$.
When $q=2-2 \cos(\nu \pi)$ with $\nu$ rational, the general 
elliptic solution degenerates into an algebraic function.
It has already been \cs{noticed}\ca{observed} \cite{paul,GB} that for the particular cases
of $q=1$, $2$ or $3$, the
resolvent obeys an algebraic equation of degree $2$, $3$, $5$ \ca{respectively}.
In this article, we will derive  from the value of $\nu$,
for any ``rational $q$'', the degree 
of the algebraic equation obeyed by the Potts-$q$
resolvent.
We will also derive the Potts-$q$ critical exponent for general values of $q$,
which agrees with Daul's \cite{daul} expression.

\newsection{The Potts-$q$ matrix model}

The Potts-$q$ model with branching interactions (which appear if one wants
 to apply the renormalization group method) is defined by the partition
 function:
\beq\label{Zdef}
Z=\int dM_1\,\dots dM_q\, \ee{-N^2  \sum_i {g\over 3 N} \tr M_i^3 
+ \psi({1\over2 N}
\tr M_i^2 ,{1 \over N} \tr \sum_{j \neq i} M_i M_j )}
\eeq
with $M_i$ hermitian matri\ca{ces} $N \times N$.

The partial derivatives of $\psi$ with respect
 to $\tr M_i^2 /(2 N)$ 
and $\tr \sum_{j \neq i} M_i M_j /N$ are
$\tilde{U}$ and $\tilde{c}/2$ respectively, and their expectation values 
(which are numbers) $U$
and $c$.
When $\tilde{U}=1$ and $\tilde{c}$ is a constant,
this reduces to the ordinary Potts-$q$
model.

We will also define the following functions:
\bea
W(z)         = & {1\over N}\moy{\tr {1\over z-M_i}}                   \cr
\td{W}(z)    = & {1\over N}\moy{\tr {1\over z-M_i}M_j}                \cr
F(z,z')      = & {1\over N}\moy{\tr {1\over z-M_i}{1\over z'-M_j}}    \cr
\td{F}(z,z') = & {1\over 2} \left[ {1\over N}\moy{\tr {1\over z-M_i}{1\over z'-M_j}M_k} + {1\over N}\moy{\tr {1\over z'-M_i}{1\over z-M_j}M_k} \right] \cr
\eea
which do not depend on the indices provided $i\neq j\neq k$.
The function $F(z,z')$ and $\td{F}(z,z')$ are thus sym\ca{m}etric:
\beq\label{Fsym}
F(z,z')=F(z',z) \qquad , \qquad \td{F}(z,z')=\td{F}(z',z)
\eeq

We will also \cs{consider} \ca{define}:
\beq
f(z)=W(z)-gz^2+(c-U)z
\eeq
Recall that $U$ and $c$ can be general functions of the numbers
 $\langle \tr M_i^2 \rangle$
and $\langle \tr M_i M_j \rangle$.

\ca{The moments $t_k$ of the resolvent $W(z)$ are defined by the large $z$ expansion:}
\cs{
We write the \cs{development}\ca{expansion} of \ca{the resolvent} $W(z)$:}
$$ W(z) \sim {1\over z} + {t_1 \over z^2} +\dots + {t_k \over z^{k+1}}
+ \dots \qquad {\rm when}\quad z\to\infty $$
and we define:
$$ u={c-U \over g} $$

\newsection{Equations of motion}

 We are going to work in the large $N$ (planar) limit,
 where we have the factorization
property: $\langle \tr A \, \tr B \rangle = \langle \tr A \rangle \langle \tr
B \rangle$ \cite{factor}.
The following changes of variables in \eq{Zdef} then 
give the following equations:

\begin{itemize}

\item{ $\delta M_1 =  {1\over z-M_1}$:}
\beq\label{eqW}
g(z^2 W(z)-z- t_1) + U (z W(z)-1) + c (q-1) \td{W}(z) = W^2(z)
\eeq

\item{ $\delta M_2 =  {1\over2}\left[ {1\over z-M_1}{1\over z'-M_2} + {1\over z'-M_2}{1\over z-M_1} \right] $:}
\bea\label{eqF}
  & g     \left( z'^2 F(z,z')-z'W(z)-\td{W}(z)  \right)  &  \cr
+ &   U    \left( z' F(z,z')-W(z)                \right)  &  \cr
+ & c     \left( zF(z,z')-W(z')                 \right)  &  \cr
+ & c (q-2) \td{F}(z,z')                                   & = W(z')F(z,z') \cr
\eea

\end{itemize}

Substracting \eq{eqF}\ with ($z\leftrightarrow z'$), and using \eq{Fsym}\ we can get rid of $\td{F}$ in \eq{eqF}:
\beq
(f(z)-f(z'))F(z,z') = g\left( z' W(z) -z W(z')+ \td{W}(z)-\td{W}(z')
 - u W(z) + u W(z') \right) 
\eeq
In particular, if we choose $z'$ such that $f(z')=f(z)$ and $z' \neq z$,
then we have:

\beq
(z'-u)W(z)-(z-u)W(z')+ \td{W}(z)-\td{W}(z') = 0
\eeq

\eq{eqW}\ allows \ca{us} to eliminate $\td{W}$, and we \ca{then} have \cs{then} an equation
 involving only $W$ or equivalently $f$:

\beq\label{eqf}
\left\{
{\begin{array}{l}
 f(z')=f(z) \hfill \cr
{1\over c}(z+z'-u-q{c\over g})f(z) = (z+z')^2-qzz'-(2-q)u(z+z')+(1-q)u^2 -{1\over c} \cr 
\end{array}}
\right.
\eeq
This seemingly difficult non-local equation is sufficient to compute
$f(z)$ \cs{entirely}.

Let us first study it perturbatively. 
When $z \rightarrow \infty$, we have $f(z) \sim -g z^2$, thus 
 $z' \sim -z$.
If we \cs{find the entire development in $z$ of $z'(z)$ and solve} \ca{expand $z'(z)$ in powers of $z$ by solving} $f(z)=f(z')$ perturbatively, then\cs{replace it} \ca{insert this expansion into} the second equation of \eq{eqf}, we obtain a set
of \cs{\ca{highly} non-trivial} equations of motion, the first of which are:
\bea
\nonumber
g t_2 +(c q-g u) t_1 & = & 0\\
\nonumber
g^2 t_4 +g (c q-2 g u) t_3 - g u (c q-2 g u) t_2 -g (g u^3+2) t_1 +c (1-q) +g u
& = & 0\\
\ldots
\eea
with $t_k = \langle \tr M_i^k /N\rangle$.
Let us recall \cs{there} that, in the general case, $u$ and $c$ are functions
of $t_2$ and $t_{1,1}=\tr M_i M_j /N$. However, we can compute $t_{1,1}$
 thanks to
\eq{eqW} in function of \cs{the matrix $M_i$ alone} \ca{the $t_k$'s}:
$$c (q-1) t_{1,1} +g t_3+U t_2 -1=0$$
Finally, our equations, contrary to ordinary equations of motion, allow \ca{us} to
 relate even traces of a given  matrix
 $M_i$ to odd traces of the same matrix \cs{exclusively}.

Only even traces appear as the higher order traces in these equations.
Let us explain why.
If we write the \ca{expansion}\cs{development} of $f(z)$ as: 
$f(z)=-g z (z-u)+1/z+\sum_{2}^{\infty} t_{i-1} /z^i$,
the $z^{-i}$ coefficient of the $f(z')-f(z)=0$ equation reads, 
as $z'=-z+u+\ldots$:
$$((-1)^i -1) \, t_{i-1}+ (-1)^{i-1}\, (i-1)\, t_{i-2} \, u + \ldots = 0$$

Thus, our loop equations give us a relation between
 the even and odd parts of
$f(z)$, and $f(z)$ can be completely determined by the requirement  that 
it has only one cut in the physical sheet (i.e. $f(-z)$ is regular along
this cut).

Finally, let us stress that these equations of motion, which are valid for
branched polymers, are a precious tool whenever one wants to study Potts-$q$
models by the renormalization group method \cite{RG}.

\newsection{Correspond\ca{e}nce with the O(n) model }

Let us \ca{now} see how to deal with a non-local equation such as \eq{eqf}.

The function $z'(z)$ defined above maps one solution of the equation 
$f(z)=y$ on another. It is involutive in the sense of multivaluated 
functions. 
We have:
$$ z'(z'(z))=z$$

Let $z_0$ be a fixed point:
$$z'(z_0)=z_0$$
and $f_0=f(z_0)$.

Then let us set
$$ \zeta=\sqrt{f_0-f} $$
and consider $z$ as a function of $\zeta$.
Then we have $z'=z(-\zeta)$ and $z(\zeta)$ is regular \ca{at}\cs{in} $\zeta=0$.
Let us set
$$ \om(\zeta)=z(\zeta)+{1\over c}{1\over 4-q}(\zeta^2-f_0-(2-q)cu) $$
\eq{eqf} rewritten in term of $\om(\zeta)$ is now \ca{an} $O(n)$-like
 quadratic equation:

\beq \label{On}
 \om^2(\zeta) + \om^2(-\zeta)+ (2-q)\om(\zeta)\om(-\zeta) = R( \zeta )
\eeq
where the right-hand side of the equation is an even polynomial of degree $4$.
The similarity between Potts-$q$ and the $O(n)$ model had already been
\ca{noted}\cs{noticed} \cite{daul,paul}, but it had always been said to be unphysical.
We \ca{shall}\cs{'ll} show here how one can relate the results for the $O(n)$ model to 
\cs{the ones}\ca{those} for Potts-$q$.

 \eq{On} can be \cs{exactly} solved \ca{exactly}, as in the case of the $O(n)$ model
 \cite{Oden}.
Here, we will
assume for simplicity that $\om(\zeta)$ has only one physical cut $[a,b]$
with $a b >0$.
Let us denote $q=2 -2 \cos(\nu \pi)$, with $0 \leq \nu \leq 1$.
Then we have, by writing  $R(\zeta+i 0)-R(\zeta-i 0)=0$:
$$(\om(\zeta+i 0)-\om(\zeta -i 0)) (\om(\zeta+i 0)+\om(\zeta -i 0)+2 \cos(\nu
\pi) \om(-\zeta))=0 \qquad \makebox{for } a \leq \zeta \leq b$$
Thus we have the linear equation:
\beq \label{lineq}
\om(\zeta +i 0)+\om(\zeta -i 0) +2 \cos(\nu \pi) \om(-\zeta) =0
\eeq
The general solution for $\om(\zeta)$ is known and can be expressed with
 elliptical functions.
 It degenerates, in the rational case (i.e. 
when $\nu$ is rational), into the solution of an algebraic equation. 
However, $\om(\zeta)$ is not the resolvent of the model as it is in the $O(n)$
model.
Indeed, it is rather the functional inverse of the resolvent for the 
Potts-$q$ model, up to some transformations.
Thus, the phase diagrams and critical exponents of Potts-$q$, as 
expected, are not the same as for the $O(n)$ model.

\newsection{Rational case}

Here we will assume that $\nu$ is rational:
$\nu=l/r$ where $l$ and $r$ are relatively prime integers.
We first recall how to obtain \eq{eqpol}, which is an
 algebraic equation for $\om(\zeta)$ \cite{Oden}. 

If we denote $\ \om_{+}(\zeta) = \ee{i \nu \pi \over 2} \om(\zeta)+
\ee{- {i \nu \pi \over 2}} \om(-\zeta) \ $ and 
$\ \om_{-}(\zeta)=\om_{+}(-\zeta)\ $,
then \eq{On} reads 
\beq \label{eqprod}
\om_{+}(\zeta) \om_{-}(\zeta) = R(\zeta)
\eeq
and \eq{lineq} becomes:
\beq \label{crosscut}
\om_{+}(\zeta+i 0)=- \ee{i \nu \pi} \om_{-}(\zeta -i 0) \qquad 
\om_{-}(\zeta+i 0)=- \ee{- i \nu \pi} \om_{+}(\zeta -i 0)
\eeq

If $\phi(\zeta)$ is defined by:
$$\om_{+}=\sqrt{R}\  \, \ee{i (\phi -{(\nu +1) \pi \over 2})} \qquad
\makebox{then we have}
 \qquad \om_{-}=\sqrt{R}\  \, \ee{-i (\phi -{(\nu +1) \pi \over 2})}$$
and
$$\om(\zeta)= - {\sqrt{R} \cos(\phi) \over \sin{\nu \pi}}$$

\eq{crosscut} shows that 
\beq \label{S}
S(\zeta)={1 \over 2} (\om_{+}^r +(-1)^{r+l} \om_{-}^r)
\eeq
has no cut, and behaves as a polynomial at infinity (as does $\om(\zeta)$),
 i.e. \ca{it} is a polynomial. We thus have the algebraic equation for $\om(z)$:
\beq \label{eqpol}
S(\zeta) = R(\zeta)^{r \over 2} \ \ee{-{i (r+l) \pi \over 2}} \
 T_r(-{\om(\zeta)\,  \sin(\nu \pi)
\over \sqrt{R(\zeta)}}) 
\eeq
where $T_r$ is the order $r$ Chebychev polynomial:
$T_r(\cos \phi)= \cos r \phi$.\\

Let us now examine the degree of this equation.
It is polynomial in $\om$ and $\zeta$, but we have to keep in mind
that the resolvent of our problem is not $\om(\zeta)$, but $W(z)=f(z)+g z (z-u)$,
with $\zeta=\sqrt{f_0 -f}$ and $\om = z -{1 \over c (4-q)}(f+(2-q) c u)$.

The right-hand side of \eq{eqpol}, seen as a polynomial in $\zeta$ and $z$,
is even and of order $2 r$ in $\zeta$, thus \ca{it is} also a polynomial in $f$.

As for the left-hand side of the equation, $S(\zeta)$ verifies $S(-\zeta)=
(-1)^{r+l} \, S(\zeta)$, thus, when $r+l$ is odd, we have to take the square 
of \eq{eqpol} to have our final algebraic equation for $f$ which should be
of degree $2 r$. When $r+l$ is even, however, \eq{eqpol} is already polynomial
in $f$, and the degree of the equation should be $r$.

Moreover, it is easy to check that the order $2 r$ terms \cs{in}\ca{on} both sides of 
\eq{eqpol} are the same.
Finally, the degree $d$ in $f(z)$ of the final equation is:
\begin{eqnarray} \label{order}
&\cr & d=2 r -1 \ \makebox{ if }\  r+l \ \makebox{ is odd} 
&\cr  &  d = \ r-1 \, \ \makebox{ if }\  r+l \ \makebox{ is even}
\end{eqnarray}
while it is of degree $d+1$ in $z$.\\
This formula generalizes the \cs{up to now}\ca{presently} known results \cite{paul,GB}:
\bea
\nonumber
\makebox{for } \nu={1/ 3}, &  q=1, & \makebox{and } d=2\\
\makebox{for } \nu={1 /  2}, & q=2, & \makebox{and } d=3\\
\nonumber
\makebox{for } \nu={2 / 3}, & q=3, & \makebox{and } d=5
\eea
In \cite{paul}, the author investigated the dilute Potts-$1$,$2$,$3$, and $4$
cases by the saddle point method. These models coincide, when there is no
dilution, with the particular case of our models with no
 branching interactions. 
This gave rise to algebraic equations when $q \neq 4$ which correspond to
our results.
However, no general result was given.
In \cite{GB}, the author used the equations of motion to solve Potts-$3$
with branching interactions, as well as for the resolution of Potts-$\infty$.
But, there again, there was no general expression.

We are now going to derive the general Potts-$q$ exponent from
\eq{On}.
If, when $\om(\zeta)$ is singular (i.e. $\zeta$ is close to one of the
bounds of the physical cut of $\om$), $\om(-\zeta)$ is not, then $\om$
cannot have more complex singularities than \ca{half-integer} exponents.
Thus, whereas in the generic case the physical cut $[a,b]$ verifies
$a b >0$, when the model is at the Potts critical point, $a$ (or $b$)
is equal to zero.
This means that the bound of the unphysical semi-infinite cut we have
suppressed by changing variables from $f$ to $\zeta=\sqrt{f_0 -f}$ coincides
\cs{then}\ca{in that case} with the physical cut.

Let us express:
$\om(\zeta)= C (-\zeta)^{\alpha} +{\rm regular\,\, part}$
\eq{On} shows immediately that the regular part is equal to zero, and that

$$\ee{2 i \pi \alpha} +1 +2 \cos (\nu \pi) \, \ee{i \pi \alpha} =0 \qquad
\makebox{i.e.} \qquad 2-2 \cos(\nu \pi) = q = 2+2 \ca{\cos} (\alpha \pi)$$
thus $$\alpha = \pm \nu + 1 +2 p \qquad \om(\zeta) \sim (f_0 -f)^{\pm \nu + 1
+2 p \over 2} \qquad p \ca{\in} Z$$
and $$f \sim (z-\cs{cste}\ca{{\rm const}})^{2 \over \pm \nu + 1 +2 p}$$
As we expect the exponent for $f$ to be greater than one, we have
the exponent ${2 \over \pm \nu + 1}$  for $f$ and the string exponent, using
\cite{kostov}'s formula, is:
$$\gamma_s=1- {2 \over \pm \nu +1}= -{(1 \pm \nu) \over (1 - \pm \nu)}$$

\newsection{Conclusion}

In this article, we \ca{have} obtained general loop equations for the
Potts-$q$ model extended to branched polymers, which
 allow \ca{us} to relate the even and odd parts of the resolvent.
This relation is then equivalent to an $O(n)$-like equation,
from which we derived the Potts critical exponents and the degree of the 
algebraic equation which appears in the rational case.
This last result generalizes \ca{the known results for $q=1$, 2 and 3 \cite{paul,GB}}\ca{ to any} 
\ca{ $q=2-2 \cos(l/r \, \pi)$ with $l<r$ integers}.
Moreover, such loop equations as we have obtained are necessary 
\ca{when one wishes} to apply the renormalization group techniques to Potts-$q$. The study
of the renormalization group flows near $q=4$ may then provide us
with useful information about the $C=1$ transition.
Since $q=\infty$ is a real $C=\infty$ model \cite{wexler} and 
the orders of the phase transitions differ between the flat and
random surface Potts models, we may find for large $q$ a set of real $C>1$ 
models.

Let us also stress that the equations of motion method may be 
used when $N$ is finite i.e. on non-planar surfaces,
 \cs{contrary}\ca{in contrast} to the saddle point method. Furthermore, they are
also less dependent on the analytic structure of the resolvent
than the saddle point method of \cite{daul,paul}.
Finally, as we know how to solve the $O(n)$ model exactly for general
values of $n$, we hope to be soon able to obtain the general expression
of the Potts-$q$ free energy and operators for any value of $q$.

\newsection{Acknowledgments}

We are grateful to R. MacKenzie and I. Kostov for careful reading of
the manuscript and F. David for useful discussions.

\end{document}